# Rutile TiO$_2$ Bulk Structural and Vibrational Properties: A DFT Study on the Importance of Pseudopotentials

Sugata Chowdhury[1], Nacole King[1], Winnie Wong-Ng[1]

[1]Materials Measurement Division, National Institute of Standards and Technology Gaithersburg, MD, 20899, USA

**Abstract:** Using first-principal density functional theory (DFT) we explained the importance of pseudopotential for decribing the electronic and vibrational properties of rutile TiO$_2$ (R-TiO$_2$). Calculations were performed using the generalized gradient approximation (GGA), local density approximation (LDA) and hybrid functional (HSE), with normconserving, ultrasoft and projector augmented wave (PAW) pseudopotential. The choice of an appropriate exchange-correlation functionals and the pseudopotential within DFT is critical to explain the different properties of R-TiO$_2$. We found that the lattice parameters predicted by the norm-conserving pseudopotential and GGA are in excellent agreement with experimental data. Our studies reveal that norm-conserving pseudopotentials with GGA exchange-correlations provide much improvement for the prediction of the electronic properties of TiO$_2$. We observed that, using LDA and ultrasoft pseudopotential is an excellent combination to describe the vibrational properties of R-TiO$_2$. Furthermore, we observed the temperature dependent asymmetric nature of vibrational modes due to changes of bond lengths and different lattice vibrations.

**Introduction:** To design an inexpensive photocatalyst for photocatalytic reactions we need materials with stable and efficient crystal structures. For these reasons transition metal-oxides have been investigated extensively for decades.[1-11] Among the various classes of photocatalysts, titanium dioxides ($TiO_2$) have attracted enormous attention due to their low cost, corrosion resistance, and surface dependent catalytic properties.[12-17] $TiO_2$ has four naturally occurring polymorphs: rutile, anatase, brookite, and an *n*-$PbO_2$ type. Electronic, vibrational and magnetic properties of rutile and anatase $TiO_2$ have been investigated extensively[14, 16-18]. The experimental band gaps of rutile and anatase $TiO_2$, are 3.0 eV and 3.2 eV, respectively with possible absorption of ultraviolet (UV) radiation.[19, 20] $TiO_2$ can be used as electrodes for photochemical reactions.[21-23] Incorporating a dopant into the $TiO_2$ systems one can usually modify the bandgap.[24-35] Non-metal doping in rutile $TiO_2$ has been one of the most investigated catalytic research areas. Doping $TiO_2$ with non-metals such as nitrogen, carbon, and phosphorus, have been reported to red-shift the band gap into the visible light region of the solar irradiance spectrum.[36, 24, 37-40] Despite extensive investigations, the fundamental physical and chemical processes behind the photocatalysis in $TiO_2$ are not totally understood. Theoretical studies would be invaluable to understand and explain the unsolved questions. Therefore, it is important to understand the different properties of $TiO_2$ using density functional theory (DFT) approaches.[14, 41-46]

Although a large amount of theoretical work regarding rutile $TiO_2$ has been published in recent years,[14, 37, 41, 44, 47-68] most of the previous benchmark works were focused on related to exchange-correlational functionals and used the generalized gradient approximation (GGA), local density approximation (LDA), hybrid Hartree-Fock/Kohn-Sham (HF/KS) methods or hybrid functional such as Heyd-Scuseria (HSE).[44-46, 69] Hybrid functional methods, such as HSE, are computationally intensive and costly. Furthermore, very few publications have discussed the effect of different pseudopotentials on the resulting properties. There is a lack of benchmark calculations regarding electronic and vibrational properties of R-$TiO_2$ and non-metal doped rutile $TiO_2$ using different pseudopotentials. In this work, we focused on the structural, electronic and vibrational properties of bulk R-$TiO_2$ using DFT to understand the effect of different exchange-correlation functionals using various pseudopotentials. The results presented here include optimized structures, densities of states, band structures and vibrational properties.

**Computational Method:** Calculations were carried out using DFT[70, 71] as implemented in Quantum Espresso (PWSCF) code.[72] Please note commercial software is identified to specify procedures only. Such identification does not imply recommendation by the National Institute of Standards and Technology (NIST). In this work, we used three different exchange-correlation functionals. Within LDA, we employed the Perdew-Zunger (PZ) exchange-correlation functionals[71] and in the case of GGA we employed the Perdew and Wang (PW),[73] PW91[74] and Perdew-Burke-Ernzerhof[75] (PBE) exchange-correlation functionals. Also, we used the HSE functional[69] to predict the band gap of $TiO_2$. We used three different pseudopotentials: norm-conserving,[76] ultrasoft[77] and projector augmented wave (PAW).[78] The kinetic energy cutoff of plane-wave expansion was taken as 520 eV. All the geometric structures were fully relaxed until the force on each atom is less than 0.002 eV/Å, and the energy-convergence criterion was $1\times10^{-6}$ eV. For the optimization structure and vibrational properties of the crystal, we used different sets of k-point grid to understand the effect of k-points convergence. In the sections of k-points convergence we explain in detail about the k-point sampling. To understand the temperature effects on vibrational properties, we considered the different lattice parameters using previously published paper by Krishna Rao et. al.[79] Here, we have optimized the atomic positions of the $TiO_2$ structure at specified temperatures and calculate the Raman modes. We compared the Raman modes of the distorted systems with the Raman modes calculated at 0 K.

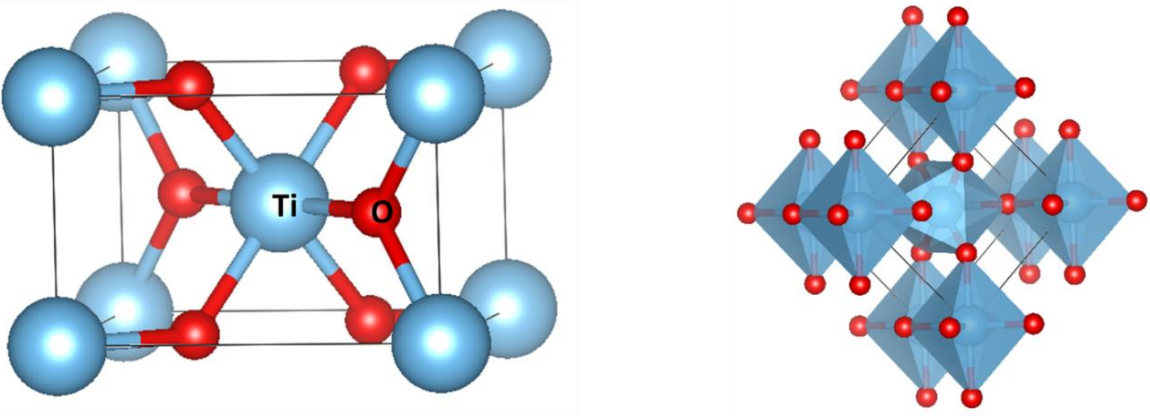

**Figure 1.** (a) Conventional unit cell of R-$TiO_2$ rutile with six atoms. (b) Coordination polyhedra of the titanium atoms. Blue spheres and octahedron represent Ti atoms, red spheres represent O atoms. The conventional unit cell outlined in solid black line.

**Initial structure model of rutile $TiO_2$:** $TiO_2$ crystalizes in four naturally occurring polymorphs: rutile, anatase, brookite, and an $n$-$PbO_2$ type. For this work, we only consider the rutile $TiO_2$ polymorph. The conventional unit cell of R-$TiO_2$, ($a$ = 4.58 Å and $c$ = 2.95 Å) is a tetragonal Bravais lattice with the $P4_2/mnm$ space group and contains six atoms. The cation ($Ti^{4+}$) is octahedrally coordinated to six anions ($O^{2-}$), and each anion is coordinated to three cations leading to an $OTi_3$ pattern as shown in Fig 1. There are three crystallographic parameters: two lattice parameters $a$, $c$ and the oxygen fractional coordinate $u$, where $u$ is the internal parameters in fractional unit. The atomic positions of Ti atoms are (0, 0, 0) and (½, ½, ½) and four oxygen atoms are at positions ($u$, $u$, 0; ½ + $u$, ½ + $u$, ½) where $u$ = 0.304937.

Ground state calculations were performed by relaxing atomic positions and lattice parameters of the structure of $TiO_2$, followed by a comparison of the lattice parameters with experimental data.

## Results

**k-point sampling**: Lattice parameters of bulk rutile $TiO_2$ were obtained in two steps. In the first step, the convergence calculations of the total energy were performed using different k-points generated with the method of Monkhorst and Pack.[80] The convergence of lattice constants and energy with respects to the k-point grid are shown in Figure 2. We used five different k-point samplings for convergence test for the unit cell and they are as follows: 4×4×4, 6×6×6, 8×8×8, 10×10×10, and 12×12×12. In Figure 2(a), the energy convergence with respect to k-point sampling is shown. Energy difference for the unit cell change was less than 0.0001 eV when we changed k-point grid from 8×8×8 to 10×10×10. This energy did not change further when we increased the k-point grid from 10×10×10 to 12×12×12. We used this procedure for all the exchange-correlations and pseudopotential. For the electronic and vibrational properties calculations, 10x10x10 and 16x16x16 k-point grids were used, respectively. In the second step, we tested the evolution of the total energy as a function of the lattice constant for R-$TiO_2$, and the results are shown in Fig 2(b). The calculations were performed using 10x10x10 k-points mesh and the theoretical equilibrium lattice constants were found. Our predicted equilibrium lattice constants using PW91 match extremely well with experimental lattice parameters and the error is 0.66 % with respect to experimental result.

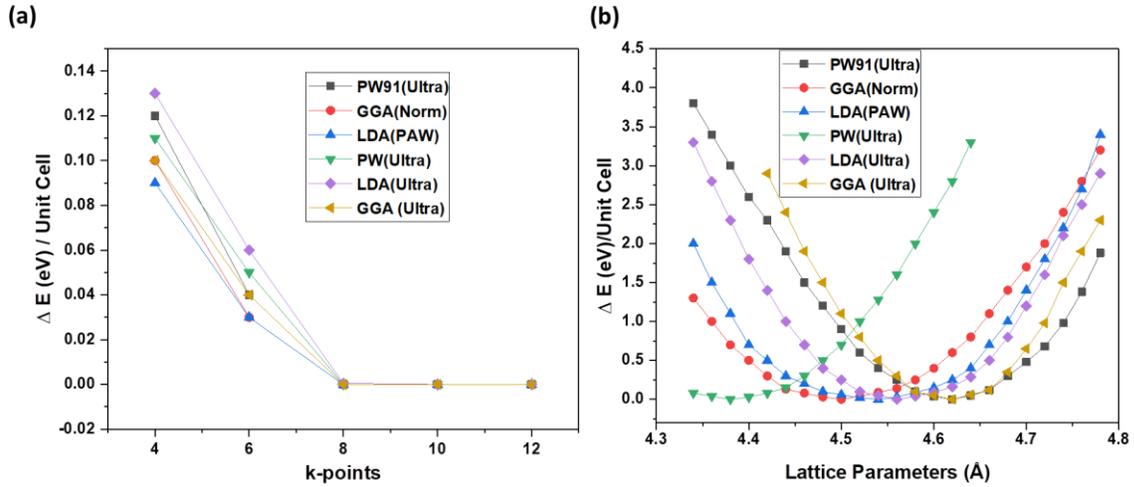

**Figure 2.** (a) Evaluation of energy with respect to the k-points. (b) Evolution of the total energy $E_{tot}$ as a function of the lattice constant rutile $TiO_2$. Pseudopotential type is described in bracket. "Norm" refers to normconserving pseudopotential and "Ultra" stands for ultrasoft pseudopotential.

**Lattice Parameters:** Relaxed structural parameters, including bond lengths are shown in Table 1. We presented six different sets of lattice parameters, Ti-O bond length and c/a ratio. Error for the computed lattice parameters using different pseudopotentials are between 0.5 % and 5 % as compared to the experimental lattice parameters. The computed lattice parameters and structural data shown in Table 1 were used in the band structure and vibrational property calculations. Lattice parameters are 0.7 % longer with PW-91 exchange-correlation in combination with norm conserving pseudopotential. Results of our calculation revealed that the errors in lattice parameters increased when we used the LDA. Ti is a 3d-transition element and is well described by the GGA functional.[81] The *c/a* ratio was also calculated, and it ranges from 0.6441 to 0.6178. The calculated internal parameters *u* is 0.3049 which is very close to the experimental value of 0.3053. The bond lengths of Ti-O are also in good agreement with the previous DFT work.[14] From the above discussion we concluded that norm conserving pseudopotential is appropriate for electronic properties prediction.

**Table 1:** Equilibrium geometry of R-TiO$_2$ calculated with different Hamiltonians, lattice parameters *a* and *c*, and Ti–O distances in Å. The pseudopotential type is described in bracket. Here "Norm" refers to the normconserving pseudopotential and "Ultra" stands for ultrasoft pseudopotential.

| Method | *a* (Å) | *c* (Å) | Ti-O (Å) | c/a |
|---|---|---|---|---|
| Experiment[82] | 4.58 | 2.95 | 1.976 | 0.6441 |
| PW91 | 4.61 | 2.97 | 1.96 | 0.6441 |
| GGA (Norm) | 4.49 | 2.89 | 1.92 | 0.6436 |
| PAW (LDA) | 4.54 | 2.89 | 1.93 | 0.6365 |
| PW | 4.37 | 2.70 | 1.85 | 0.6178 |
| LDA (Ultra) | 4.55 | 2.88 | 1.94 | 0.6329 |
| GGA (Ultra) | 4.63 | 2.98 | 1.97 | 0.6328 |

**Electronic Band Structure**: The electronic band structures of the bulk R-TiO$_2$, which were computed using different exchange correlations and pseudopotentials, are shown in Figure 3. The calculated direct and indirect bandgaps of R-TiO$_2$ are also given in Figure 3. The calculated band gap is underestimated with respect to the experimental value regardless of the exchange-correlations or pseudopotentials used. The experimentally reported bandgap of rutile is 3.0 eV; in our computational calculations the band gaps are between 1.87 eV to 2,53 eV. The band gap value using GGA exchange correlations and norm-conserving pseudopotential is 2.53 eV, which is a much better value than the previously reported data using GGA or LDA. Our calculated bandgap is 3.19 eV when using the HSE functional and is in excellent agreement with experimental findings. From our electronic band structure calculations, it is clear that the conduction band minimum at Γ-point is almost the same as the conduction band minima at point R and M. This degeneracy is visible for all of exchange-correlations and pseudopotentials. We concluded that norm-conserving pseudopotential predicted the band gap better as compared to ultrasoft and PAW.

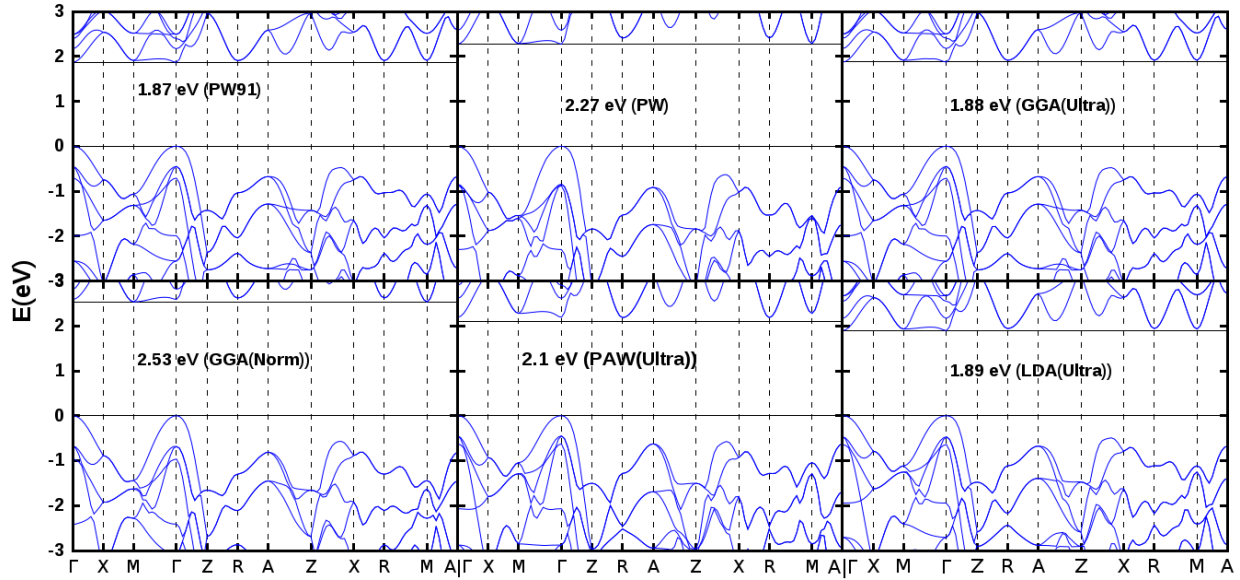

**Figure 3.** Electronic band structures of R-TiO$_2$, using different exchange-correlations and pseudopotentials along the high symmetry lines of the Brillouin zone. Pseudopotential type is described in brackets.

**Projected density of states (PDOS):** The total and partial densities of states (PDOS) are depicted in Figure 4. Our calculations revealed that the 2p orbital of O atom and 3d orbital of Ti atom play a major role for formation of any type of sub-band and the hybridization process. From the PDOS of R-TiO$_2$, it is evident that the conduction bands are mainly emerging from the 3d orbital of Ti atoms and the valence bands are dominated by the 2p orbital of O atoms, whereas, the contribution of the O$_{2s}$ orbital at the Fermi level is negligible. From our calculations we found non-zero Ti-PDOS in the valance band and non-zero O-PDOS in the conduction band around the Fermi level. These nonzero PDOS indicates the strong hybridization between the 3d orbitals of Ti atoms and the 2p orbital of O atoms. Also, we found two distinct peaks in the PDOS of Ti$_{3d}$ orbital indicating a distinct separation into sub-bands. The crystal-field theory of transition-metal (di)oxide complexes explain the sub-bands separation of the 3d orbital of Ti atoms.[83] According to crystal field theory, those *d*-states are split into the twofold-degenerate $e_g$-like states with $d_{x2-y2}$ and $d_{z2}$ character and energetically lower lying threefold degenerate $t_{2g}$-like $d_{xy}$, $d_{yz}$ and $d_{xz}$ types states. It is also evident that the upper valence bands are emanating due to hybridization between the 3d bands of Ti-ion and 2p band of O-ions in $OTi_3$. This hybridization leads to a formation of bandwidth of 5.75 eV to 6 eV, approximately in the upper states of the valance band. The width of the upper valence has been reported to be 5.4 eV by Manghnani *et. al.*[84, 85] Our calculations revealed that a combination of GGA and norm-conserving pseudopotential is the best way to describe the electronic structure of $TiO_2$ other than the HSE functional.

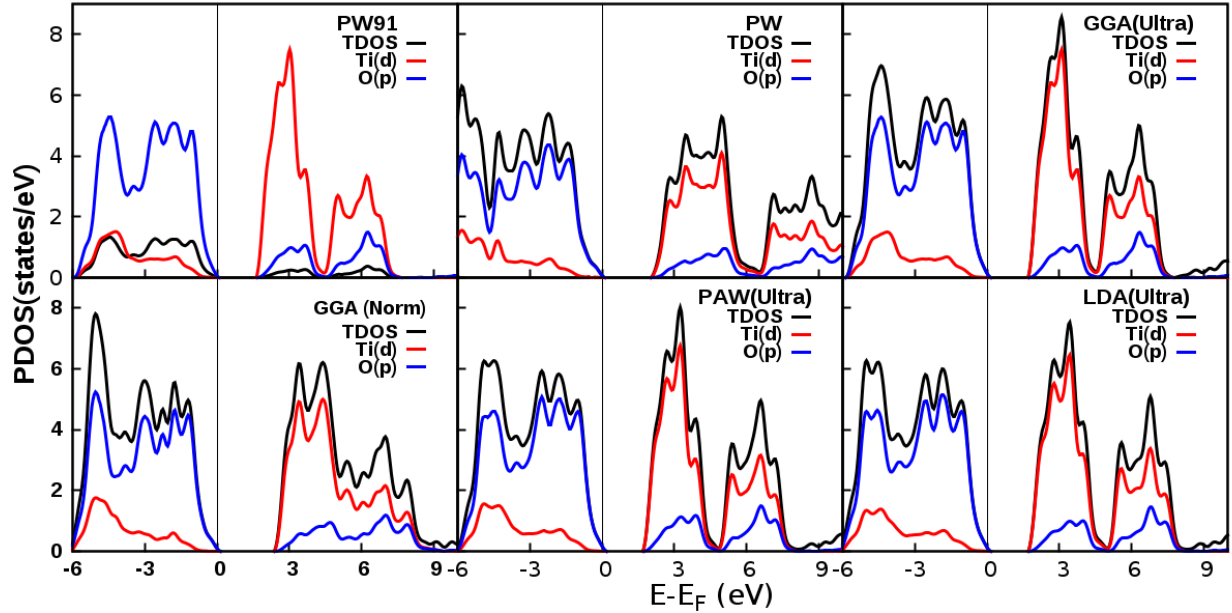

**Figure 4.** The total and the partial density of states (DOS) of R-TiO$_2$, as obtained using different exchange-correlation functionals. The black lines denote the total DOS; the partial DOS of titanium and oxygen are represented in blue and red lines, respectively. The Fermi level has been arbitrarily chosen as the origin of the energy. The vertical, black line indicates the position of the Fermi energy (EF) which has been set equal to zero. The left-hand side box of each diagram illustrates the valence band feature, and the right-hand side the conduction band. Pseudopotential type is described in brackets.

**Vibrational Properties**: In this section, the vibrational properties of R-$TiO_2$, specifically the Raman-active phonon modes are discussed. The vibrational modes of R-$TiO_2$ at the Γ point can be described by the following irreducible representations:

$$\Gamma_{opt} = A_{1g} + A_{2g} + A_{2u} + 2B_{1u} + B_{1g} + B_{2G} + E_g + 3_{Eu}$$

where the symbols g represents Raman active mode, u represents the infrared active, whereas E represents degenerate mode. In Figure 5, the Raman-active modes of rutile $TiO_2$ are shown where the $A_{1g}$ and $B_{1g}$ are the in-plane modes, the E**g** mode is the out-of-plane mode, and the $B_{2g}$ mode is the breathing mode. All these $B_{1g}$, $E_g$ and $A_{1g}$ modes describe the motions of the oxygen ions. In Table 2, we compare the Raman-active phonon modes with experimental findings. The high frequency Raman modes ($A_{1g}$ and $B_{2g}$) calculated using LDA exchange-correlations functional in combination with ultrasoft pseudopotential agree very well with experimental results,[86] whereas low frequency Raman mode ($B_{1g}$) also matches very well when we use GGA exchange-correlational functional in combination with ultrasoft pseudopotential. The active Raman mode predictions given by other exchange-correlations functional deviate significantly from experimental findings. From above discussion we conclude that ultrasoft pseudopotential predicted the position of Raman modes much better than any other pseudopotentials.

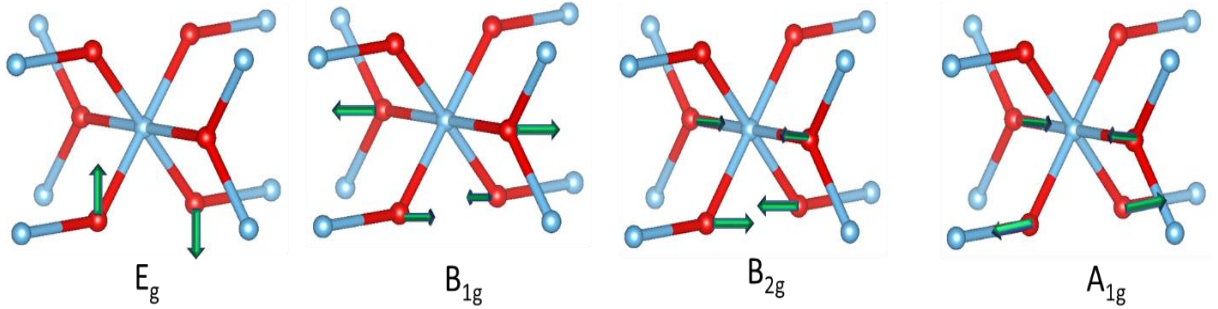

**Figure 5**. Raman active modes of R-TiO$_2$. Blue balls represent Ti and red balls O atoms. The green color arrow represents the eigenvector of the Raman modes.

**Table 2**. The active Raman-modes calculated using different exchange correlations and pseudopotential. Pseudopotential type is described in bracket.

| Method | B$_{1g}$ (cm$^{-1}$) | E$_g$ (cm$^{-1}$) | A$_{1g}$ (cm$^{-1}$) | B$_{2g}$ (cm$^{-1}$) |
|---|---|---|---|---|
| Experimental[86] | 143 | 447 | 612 | 826 |
| PW91 | 139.7 | 436.3 | 580.3 | 779.5 |
| GGA (Norm) | 156.1 | 465.6 | 645.2 | 738.5 |
| PAW (LDA) | 124.8 | 464.8 | 614.0 | 818.6 |
| PW | 121.9 | 454.4 | 587.3 | 775.7 |
| LDA (Ultra) | 125.8 | 469.8 | 615.0 | 823.7 |
| GGA (Ultra) | 146.0 | 432.4 | 572.6 | 778.2 |

We also evaluated the Raman active phonon modes as a function of temperature using the LDA exchange-correlations functional and ultrasoft pseudopotential, as they predict very well the positions of Raman modes. As mentioned in the computational section earlier, to understand the temperature effect on vibrational properties we considered different lattice parameters using previously published paper by Krishna Rao et. al.[79] The Raman modes of distorted systems with respect to the phonon dispersion calculated at 0 K were compared. In Fig. 5, we present the variation of Raman modes as a function of temperature. We found the softening of the out-of-plane $E_g$ modes while the $B_{1g}$ modes harden with an increase of temperature. It is also observed that both the $A_{1g}$ and the $B_{2g}$ modes soften slightly when the temperature increases. Our results are in excellent agreement with previous reports.[87, 88]

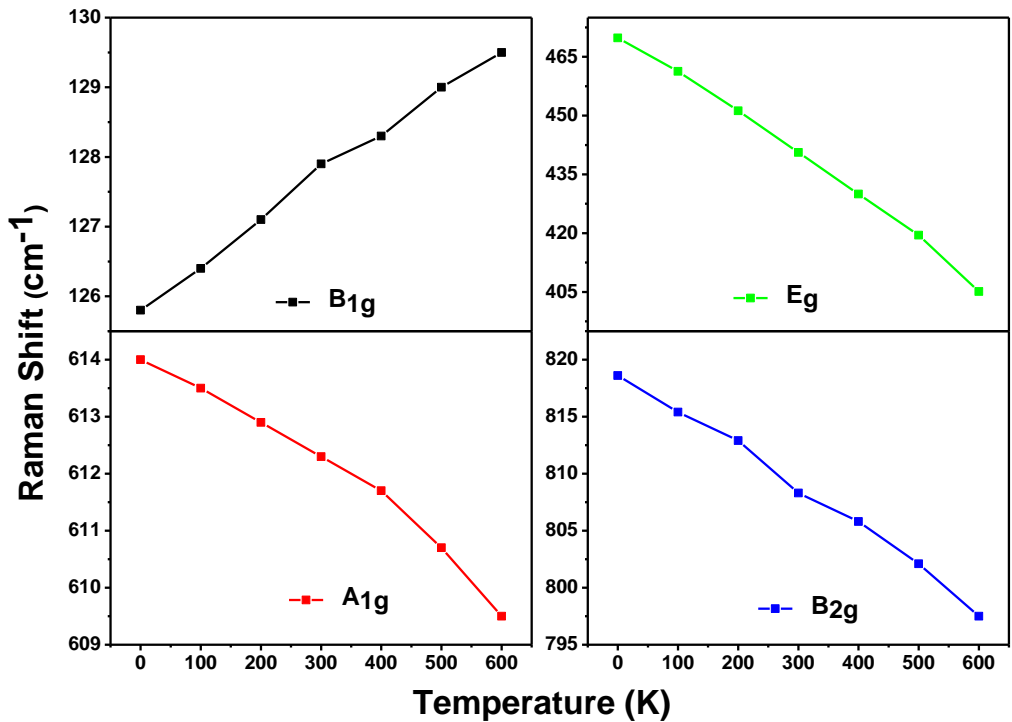

**Fig 6**. The active Raman-modes calculated using LDA (ultrasoft) at different temperature.

From Fig 6. It is evident that the high energy vibrational modes of R-TiO$_2$ decrease as the temperature increases, whereas the lower energy vibrational mode ($B_{1g}$) increases as temperature increases. We conclude that different lattice vibrations respond differently in R-TiO$_2$. As the temperature increases the Ti-O bonds become weaker and the relaxation time for the higher energy vibrational modes decreases. Response of different lattice vibrations indicates the asymmetric lattice vibrational characteristics of rutile $TiO_2$.

**Conclusion:** This work provides us with an understanding of the effects of pseudopotentials and exchange correlations on the properties of R-$TiO_2$. We have investigated the electronic and vibrational properties of R-$TiO_2$. We found that the best calculated lattice parameters and bond length of R-$TiO_2$ is obtained by PW91 exchange-correlations in combination with the norm-conserving pseudopotential (within 1 % from the experimental results). The electronic properties of rutile $TiO_2$ can be described very well by GGA exchange-correlations in combination with norm-conserving pseudopotentials. The present calculations also show the positions of Raman modes varying with the type of pseudopotentials. The positions of the active Raman-modes obtained by LDA are in very good agreement with reported experimental literature. We also explained the positions of Raman modes as a function of temperature and described the asymmetric response of the vibrational modes of rutile $TiO_2$. Based on these results, normconserving pseudopotential appears to be a good compromise to obtain an accurate description of structural and electronic properties, and ultrasoft pseudopotential is an excellent choice to describe the vibrational properties of rutile $TiO_2$. It is anticipated that our approach and results presented in this work will help us predict and explain the electronics and viabrational properties of other oxides as well.

**Acknowledgement:** We thank Dr. Kamal Choudhary and Dr. Eric Cockayne at NIST for helpful discussions. S.C. performed all DFT calculations and worked on data analysis and verification. N.K. and W.W-N assisted in writing the manuscript and providing scientific discussions. All the DFT calculations were done in using NIST supercomputing center.